\documentclass[aps,prd,preprint]{revtex4-1}
\usepackage{amsmath,amssymb,graphics,graphicx,color}
\usepackage{hyperref}

\begin{document}
	
\title{Higgs Inflation With Four-form Couplings }
\author{C.J. Ouseph$^{a,b}$, Kingman Cheung$^{a,b,c,d}$ }
\affiliation{
  $^a$ Department of Physics, National Tsing Hua University, Hsinchu 300,
  Taiwan\\
  $^b$ Institute of Astronomy ,National Tsing Hua University, Hsinchu 300,
  Taiwan\\
  $^c$ Physics Division, National Center for Theoretical Sciences,
  Hsinchu 300, Taiwan \\
  $^d$ Division of Quantum Phases and Devices, School of Physics,
  Konkuk University, Seoul 143-701, Republic of Korea
}
\date{\today}

\begin{abstract}
  We consider a new inflationary model in which an antisymmetric
  tensor field $A_{\nu\rho\sigma}$ and its four-form field strength
  $F_{\mu\nu\rho\sigma}=4\partial_{[\mu} A_{\nu\rho\sigma]}$ are
  coupled to the scalar sector of the standard model and to the Ricci
  scalar $\mathcal{R}$. The four-form field induces modifications to
  the Higgs self-coupling constant, the cosmological constant, and the
  non-minimal coupling constant, which results in the modification to
  the inflaton potential. We also show that there is no need for the
  Higgs-gravity coupling in the presence of four-form-gravity
  interaction, but still can produce the right amount of density
  perturbation for inflation.
\end{abstract}

\maketitle

\section{Introduction}

Recently, an interesting proposal was put forward to give a simultaneous
solution to the smallness of both cosmological constant and the Higgs vacuum
expectation value (VEV) \cite{b,c}. This is based on the argument that
if the Higgs VEV is explained by a choice in cosmological selection among
the landscape, the cosmological-constant scale should also be addressed
by the same selection.

The framework is the standard model (SM) plus a non-dynamical field
described by a {\it 4-form}.  The 4-form can arise from some nontrivial topological sectors.
When the
4-form couples to gravity, it contributes to the vacuum energy. If 
it couples to the Higgs field at the same time, it can generate a
number of configurations for the Higgs VEV, so that the acceptable
Higgs-boson VEV and mass can be selected.  Simultaneously, the
smallness of the Cosmological Constant can be selected.
The coupling of the 4-form to gravity was also proposed to address
the problem of inflation \cite{4-inflation}, and more recently in
Refs.~\cite{d,e}.

Cosmological inflation is the most favorable theory of early universe.
It not only explains the absence of a number of relics that should
have existed from the Big Bang, but also provides the seeds
for the growth of structures in the Universe.
In the last two decades, people have been attempting to figure out
the most promising candidate for cosmological inflation. There are
plenty of attempts to address inflation within the
framework of the standard model (SM) and theories beyond the SM.
The SM Higgs field is always a fascinating candidate as the
inflaton because of the non-requirement of additional scalar degree of
freedom. However, the minimal Higgs inflation model is not favorable,
if not ruled out, because of the fine-tuned value of the Higgs
self-coupling constant $\lambda$.
A non-minimal coupling between the SM Higgs field and the Ricci
scalar $\mathcal{R}$ \cite{a} was then introduced with the hope of
relaxing the value of $\lambda$.
However, such kind of attempts may lead to unitarity violation.
In order to avoid the problem of unitary violation in Higgs
sector, we study the scenario that
the four-form field $F_{\mu\nu\rho\sigma}$ is coupled to the Higgs
field and the Ricci scalar.  Consequently, the Higgs self-coupling
constant and the non-minimal coupling constant are modified.
In addition to the cosmological constant $\Lambda$ being
scanned to an effective value $\Lambda_{eff}$, the Higgs-boson
mass $m_{gen}$ can also be generated before the electroweak symmetry
breaking \cite{b,c}.

In this work, we consider two different inflationary models of
non-minimal couplings. The first one is the more general model of
inflation, in which both the SM Higgs and the four-form fields couple
to gravity and it results in an effective 
non-minimal coupling constant $\xi_{eff}$ in the model.
In the second  model, only the non-minimal coupling of $F_{\mu\nu\rho\sigma}$
with gravity is considered. The SM Higgs is free from the
non-minimal coupling, because of non-necessity of
Higgs-gravity coupling for inflation.
We then compare the results of the minimal and non-minimal Higgs inflation
models, with respect to the most recent values of spectral index and
the tensor-to-scalar ratio.

The organization is as follows.
In Sec.~II, we describe the non-minimal coupling models. In Sec.~III, we
calculate the spectral index and tensor-to-scalar ratio of the models.
Section III A and B deal with the comparison between the minimal
and non-minimal coupling models. We conclude in Sec. IV.

\section{The 4-Form Interactions}

Let us first consider an inflationary model in which the SM Higgs field is
coupled to the scalar curvature $\mathcal{R}$, to  the tensor 3-form field
$A_{\nu\rho\sigma}$, and to its four-form field strength
$F_{\mu\nu\rho\sigma}=4\partial_{[\mu}   A_{\nu\rho\sigma]}$ 
in a non-minimal way.
Here we are writing the Higgs field $H=h/\sqrt{2}$ in the unitary gauge
without specifying the term with the VEV.
The Lagrangian for the Higgs field $h$ is given by
\begin{equation}
\mathcal{L}=\mathcal{L}_0+\mathcal{L}_{int}+\mathcal{L}_s+\mathcal{L}_L+\mathcal{L}_{memb}
\end{equation}
where
\begin{eqnarray}
  \mathcal{L}_0=\sqrt{-g}\big[M^2_{\rm PL}\frac{\mathcal{R}}{2}+
    \frac{b^2\mathcal{R}^2}{2}-\frac{1}{48}F^{{\mu\nu\rho\sigma}}F_{\mu\nu\rho\sigma}
    -
    \frac{1}{2}(\partial_{\mu}h)^2 -
    \frac{\lambda}{4}h^4  - \Lambda\big]\\
\mathcal{L}_{int}=\frac{C_1}{48}\epsilon^{\mu\nu\rho\sigma}F_{\mu\nu\rho\sigma}h^2+\frac{1}{2}\xi_{1}h^2\mathcal{R}+\xi_{2}\epsilon^{\mu\nu\rho\sigma}F_{\mu\nu\rho\sigma}\mathcal{R}\\
\mathcal{L}_{s}=\frac{1}{6}\partial_{\mu}\big[\big(\sqrt{-g}F^{{\mu\nu\rho\sigma}}-C_1\epsilon^{\mu\nu\rho\sigma}h^2+\epsilon^{\mu\nu\rho\sigma}\xi_{2}\mathcal{R}\big)A_{\nu\rho\sigma}\big]\\
\mathcal{L}_{L}=\frac{q_1}{24}\epsilon^{\mu\nu\rho\sigma}\big[F_{\mu\nu\rho\sigma}-4\partial_{[\mu} A_{\nu\rho\sigma]}\big]\\
\mathcal{L}_{memb}=\frac{q_2}{2}\int d^3\zeta\big[\delta^4(x-x(\zeta))A_{\nu\rho\sigma}\frac{\partial x^\nu}{\partial\zeta^a}\frac{\partial x^\rho}{\partial\zeta^b}\frac{\partial x^\sigma}{\partial\zeta^c}\epsilon^{abc}     \big]
\end{eqnarray}
We are considering the same form of Lagrangian as in Refs.~\cite{d,e}.
We have introduced the terms
$\xi_{1}h^2\mathcal{R}$ and
$\xi_{2}\epsilon^{\mu\nu\rho\sigma}F_{\mu\nu\rho\sigma}\mathcal{R}$
in the $\mathcal{L}_{int}$, where 
$\xi_{1}$ is the non-minimal coupling of Higgs field with gravity
and the four-form field is coupled to the Ricci scalar via
$\xi_{2}$. The $\Lambda$ is the cosmological constant.
The required Lagrangian for the inflationary scenario
obtained after integrating out the $F_{\mu\nu\rho\sigma}$ field
can be rewritten as,
\begin{equation}
\begin{aligned}
  \mathcal{L}=\sqrt{-g}\big[M^2_{\rm PL}\frac{\mathcal{R}}{2}+\frac{b^2\mathcal{R}^2}{2}-\frac{1}{2}(\partial_{\mu}h)^2-\frac{\lambda}{4}h^4
    -  \Lambda+ \\
    \frac{1}{2}\xi_{1}h^2\mathcal{R}-\frac{1}{2}(C_1h^2-\xi_{2}\mathcal{R}+q_1)^2-\frac{1}{6}\epsilon^{\mu\nu\rho\sigma}\partial_{\mu}q_1A_{\nu\rho\sigma}+\mathcal{L}_{memb}\big] \;.
\end{aligned}
\end{equation}\\
From the equation of motion of $A_{\nu\rho\sigma}$, 
\begin{equation}\epsilon^{\mu\nu\rho\sigma}\partial_{\mu}q_1=\frac{q_2}{2}\int d^3\zeta\big[\delta^4(x-x(\zeta))\frac{\partial x^\nu}{\partial\zeta^a}\frac{\partial x^\rho}{\partial\zeta^b}\frac{\partial x^\sigma}{\partial\zeta^c}\epsilon^{abc}     \big]
\end{equation}
thus $q_1$ get quantized by $q_1=nq_2$, where $n$ is an integer.


Other than the non-minimal couplings of Higgs field with gravity
and the 4-form field with the gravity, there is also
  the $\mathcal{R}^2$ coupling.
In order to simplify the Lagrangian we perform a dual transformation
of the $\mathcal{R}^2$ term in Eq.~(7) in terms of a real scalar $\psi$,
$\psi=\mathcal{R}\sqrt{b^2-\xi_{2}^2}\times\frac{(1\pm i\sqrt{3})}{2}$,
and we obtain
\begin{equation}
  \label{eq9}
  \mathcal{L}=\sqrt{-g}\big[\frac{\Omega(h,\psi,q)}{2}\mathcal{R}
    -  \frac{1}{2}(\partial_{\mu}h)^2-\Lambda-\frac{q_1^2}{2}-h^4
    (\frac{\lambda}{4}+\frac{C_1^2}{2})-\frac{1}{2}h^2(2q_1c_1)
    -\frac{1}{2}\psi^2        \big]
\end{equation}  
with 
\begin{equation}
\Omega(h,\psi,q)=\Big[(M^2_{\rm PL}+h^2(\xi_{1}+2C_1\xi_{2})+2\xi_{2} q_1)+\sqrt{b^2-\xi_{2}^2}\psi\Big] \;.
\end{equation}
Here we make a field redefinition:
\begin{equation}
\sigma=(\xi_{1}+2C_1\xi_{2})\frac{h^2}{\xi_{2}}+2q_1+\frac{\sqrt{b^2-\xi_{2}^2}}{\xi_{2}}\psi
\end{equation}
and Eq.~(\ref{eq9}) then becomes
\begin{equation}
  \label{eq12}
\begin{aligned}
  \mathcal{L}=\sqrt{-g}\big[   \frac{1}{2}(M^2_{\rm PL}+\xi_{2}\sigma)\mathcal{R}
    -
    \frac{1}{2}(\partial_{\mu}h)^2-\Lambda-\frac{q_1^2}{2}- \\
    h^4(\frac{\lambda}{4}+\frac{C_1^2}{2})-\frac{1}{2}h^2(2q_1c_1)-\frac{1}{2}\frac{\xi_{2}^2}{{b^2-\xi_{2}^2}}(\sigma-(\frac{h^2}{\xi_{2}}(\xi_{1}+2\xi_{2} C_1)+2q_1))^2      \big] \;.
\end{aligned}
\end{equation}
The potential of the Lagrangian can be expanded as
\begin{equation}
\begin{aligned}
\noindent V(h,q,\sigma)=h^4\lambda_{eff} +m^2_{gen}h^2+\Lambda+\frac{q_1^2}{2}(1-4\beta)-2q_1\beta\sigma+\beta\frac{\sigma^2}{2} \;,
\end{aligned}
\end{equation}
where $\beta=\frac{\xi_{2}^2}{{b^2-\xi_{2}^2}}$,
and $\lambda_{eff}$ defined as,
\[
\lambda_{eff}=\Big(\frac{\lambda}{4}+\frac{C_1^2}{2}+\frac{\beta}{2}(\frac{\xi_{1}}{\xi_{2}^2}+\frac{2C_1}{\xi_{2}})^2\Big) \;.
\]
The modified Cosmological Constant becomes
\[
\Lambda_{\rm eff}=\Lambda+\frac{q_1^2}{2}(1-4\beta) \;.
\]
The generated Higgs mass from the 4-form interaction can be obtained as
\[
 m^2_{\rm gen}=h^2\Big[c_1q_1+\beta\Big(\frac{\xi_{1}}{\xi_{2}^2}+\frac{2C_1}{\xi_{2}}\Big)(2q_1-\sigma)      \Big]\;.\]
It was shown that desirable values for Higgs mass and cosmological constant
can be achieved with scanning \cite{b,c}.
   
\section{Cosmological Inflation and Density Fluctuation}

A lot of works in literature have discussed about the SM Higgs field
as the inflaton. Here we introduce a new scenario in which
we describe how the Higgs-gravity coupling and the coupling
between Higgs and the 4-form flux can give rise to successful inflation.
The Lagrangian in Jordan frame can be described as follows   
\begin{equation}
\mathcal{S}_J=\int d^4x\sqrt{-g}\Big(\frac{1}{2}(M^2_{\rm PL}+\xi_{2}\sigma)\mathcal{R}+\frac{1}{2}(\partial_{\mu}h)^2- V(h,q_1,\sigma)\Big) \;.
\end{equation}
We can integrate out the $\sigma$ field from the Eq~(\ref{eq12})
$\sigma=(\xi_{1}+2C_1\xi_{2})\frac{h^2}{\xi_{2}}+2q_1$.
However, some components of the the generated Higgs mass $m_{gen}$ and the
$\lambda_{eff}$ disappear after
integrating out the $\sigma$ field from the Lagrangian.
\begin{eqnarray*}
\lambda_{eff}=\frac{\lambda}{4}+\frac{C_1^2}{2}\\
 m^2_{\rm gen}=h^2c_1q_1 \;.
\end{eqnarray*}
In this inflation scenario, we only consider 
the highest power of the Higgs field, the $h^4$, as the inflaton
in the potential, i.e., we are neglecting the $h^2$ term.
We can get rid
of the non-minimal coupling of gravity from the Lagrangian using
the conformal transformation \cite{f}.  Here we are performing the conformal
transformation from the Jordan frame to the Einstein frame:
\begin{equation}
  \hat{g}_{\mu\nu}=\Omega g_{\mu\nu}, \;\;
  \Omega=(1+\frac{h^2\xi_{eff}+2\xi_{2} q_1}{M^2_{\rm PL}}),
  \;\; \rm{with} \;\;
  \xi_{eff}=\xi_{1}+2C_1\xi_{2}\;.
\end{equation}
The conformal transformation gives rise to a non-minimal kinetic
term for the inflaton field.  We redefined the field as 
\begin{equation}
\frac{d\phi}{dh}=M_{\rm PL}\sqrt{\frac{\Omega}{\Omega^2}+\frac{3}{2}\frac{(\frac{d\Omega}{dh})^2}{\Omega^2}}
\end{equation}
with
\begin{equation}
\phi=\sqrt{\frac{3}{2}}M_{\rm PL}\ln\Omega \;.
\end{equation}
The action in the Einstein frame is defined as 
\begin{equation}
\mathcal{S}_E=\int d^4x\sqrt{-\hat{g}}\Big(\frac{M^2_{\rm PL}\mathcal{\hat{R}}}{2}+\frac{1}{2}\partial_{\mu}\phi\partial^{\mu}\phi-U(\phi)\Big)
\end{equation}
where $U(\phi)=\frac{V(h,q_1)}{\Omega^2}$ with $V(h,q_1)=\lambda_{eff}h^4$. 
The flat exponential inflaton potential can be expressed as
\begin{equation}
U(\phi)=\frac{M^4_{ \rm PL}\lambda_{eff}}{\xi_{eff}^2}\Big[\frac{\big(\exp(\sqrt{\frac{2}{3}}\frac{\phi}{M_{\rm PL}})-(1+\frac{2\xi_{2} q_1}{M^2_{\rm PL}})\big)^2}{\exp(2\sqrt{\frac{2}{3}}\frac{\phi}{M_{\rm PL}})}\Big]
\end{equation} with $\lambda_{eff}=\frac{\lambda}{4}+\frac{C_1^2}{2}$.

The slow-roll parameters can be calculated analytically as a function
of $h(\phi)$:
\begin{eqnarray}
  \epsilon&=&\frac{M^2_{\rm PL}}{2}\Bigg(\frac{\frac{\partial U}{\partial\phi}}{U}
  \Bigg)^2=\frac{4(M^2_{\rm PL}+2\xi_{2} q_1)^2}{3\xi_{eff}^2h^4}\\
  \eta&=&M^2_{\rm PL}\Bigg(\frac{\frac{\partial^2U}{\partial\phi^2}}{U}\Bigg)=
  \Big(\frac{4}{3}\Big)\frac{(M^2_{\rm PL}+2\xi_{2} q_1)(M^2_{\rm PL}+2\xi_{2} q_1-\xi_{eff}h^2)}{\xi_{eff}^2h^4}
\end{eqnarray}
Note that $\epsilon=1$ at the end of slow roll.
The field value of inflaton at the end of inflation is given by 
$h_{end}=\big(\frac{4}{3}\big)^{\frac{1}{4}}M_{\rm PL}
\sqrt{\frac{(1+\frac{2\xi_{2} q_1}{M^2_{\rm PL}})}{\xi_{eff}}}$.
The number of $e$-foldings can be calculated as
\begin{equation}
  \mathcal{N}=\int_{h_{end}}^{h_0}\frac{1}{M_{\rm PL}^2}
  \Bigg(\frac{U}{\frac{\partial U}{\partial h}}
  (\frac{\partial\phi}{\partial h})^2\Bigg) dh \;.
\end{equation}
For $\mathcal{N}=60$ the value of $h_0=9.00859\times M_{\rm PL}
\sqrt{\frac{(1+\frac{2\xi_{2} q_1}{M^2_{\rm PL}})}{\xi_{eff}}}$.
We can constrain the value of $\lambda_{eff}$ and $\xi_{eff}$ with the
COBE normalization $\frac{U}{\epsilon}=(0.027M_{\rm PL})^4$.
Now the value of $\frac{\lambda_{eff}}{\xi_{eff}^2}=1.10256\times10^{-10}$.
The spectral index $n_s=1-6\epsilon+2\eta$ and the tensor-to-scalar
ratio $r=16\epsilon$ for $\mathcal{N}=60$ is calculated to be
$n_s=0.9633$, $r=0.0032$.
The predicted values for $n_S$ and $r$ are in excellent
agreement with the
Planck 2018 data within $1\sigma$ \cite{g}.
The reheating era started at the end of the slow roll. The SM Higgs field
would start to interact with other SM particles at the time of reheating,
$T_{reh}\simeq(\frac{2\lambda_{eff}}{\pi^2g^*})^{\frac{1}{4}}
\frac{M_{\rm PL}}{\sqrt{\xi_{eff}}}\simeq 2\times10^{15}$ GeV,
where $g^*$ is the number of degrees of freedom of the SM at
the time of reheating and $g^*=106.75$.

\subsection{Realization of Non-Minimal Higgs Inflation with $\xi_{2}=0$ limit}

By switching off $\xi_2$ ($\xi_{2}=0$) we can reproduce the
same form of inflaton potential of Bezrukov and Shaposhnikov's \cite{a} with
$\lambda_{eff}=\frac{\lambda}{4}+\frac{C_1^2}{2}$
(by switching off $C_1$ ($C_1=0$), the entire model become the
non-minimal Higgs inflation):
\begin{equation}
U_E=\frac{\lambda_{eff} M^4_{\rm PL}}{\xi^2}\Big(1-\exp\big(-\frac{2\phi}{\sqrt{6}M_{\rm PL}}\big)\Big)^2
\end{equation}
where $\xi=\xi_{1}$, the Higgs-gravity coupling. The slow-roll parameters
then become the same form as those of the above model:
\begin{eqnarray}
\epsilon= \frac{4M^2_{\rm PL}}{3\xi_{1}^2h^4},\\
\eta= -\frac{4M^2_{\rm PL}}{3\xi_{1} h^2}\;.
\end{eqnarray}
From the COBE normalization
$\frac{\lambda_{eff}}{\xi_{1}^2}\sim10^{-10}$.  This model was
arguing the requirement of non-minimal coupling ($\xi_{1}$) of Higgs
sector with the curvature tensor. The coupling of SM Higgs with the
four-form plays a crucial role in this scenario as well. This
model [$\lambda_{eff}=\frac{\lambda}{4}$] of inflation is questionable
because it may violate unitarity before reaching the inflationary
scale\cite{h,i,j,k}. If we address the inflation with an effective self-coupling
constant, it is possible to reduce the value of non-minimal coupling
constant.

\subsection{Minimal Higgs Inflation with the Non-Minimal Coupling
  of Tensor Field with gravity }

We have the freedom to choose different combinations of couplings
with the gravity.
If we switch  off $\xi_1$ ($\xi_{1}=0$), then
$\xi_{eff}=2C_1\xi_{2}$. We can then obtain another inflationary scenario
from this kind of coupling without changing the values of any other
inflation parameters. The inflaton potential is given by 
\begin{equation}
U(\phi)=\frac{M^4_{\rm PL}\lambda_{eff}}{(2C_1\xi_{2})^2}\Big[\frac{\big(\exp(\sqrt{\frac{2}{3}}\frac{\phi}{M_{\rm PL}})-(1+\frac{2\xi_{2} q_1}{M^2_{\rm PL}})\big)^2}{\exp(2\sqrt{\frac{2}{3}}\frac{\phi}{M_{\rm PL}})}\Big] \;,
\end{equation}
where $\lambda_{eff}=\Big(\frac{\lambda}{4}+\frac{C_1^2}{2}\Big)$.
The slow-roll parameters for this scenario can be obtained
\begin{eqnarray}
\epsilon&=&\frac{4(M^2_{\rm PL}+2\xi_{2} q_1)^2}{3(2C_1\xi_{2})^2h^4}\\
\eta&=&\Big(\frac{4}{3}\Big)\frac{(M^2_{\rm PL}+2\xi_{2} q_1)(M^2_{\rm PL}+
  2\xi_{2} q_1-(2C_1\xi_{2})h^2)}{(2C_1\xi_{2})^2h^4} \;.
\end{eqnarray}
The values for $n_s$ and $r$ can be calculated using the
formalism that we discussed in the previous subsection. They preserve their
values because of the form of the potential. For $\mathcal{N}=60$ the
calculated values for $n_s$ and $r$ equal $n_s=0.9633$ and $r=0.0032$ with
$\lambda_{eff}=\Big(\frac{\lambda}{4}+\frac{C_1^2}{2}\Big)$ and
$\xi_{eff}=2C_1\xi_{2}$. Using the COBE results
$\frac{\lambda_{eff}}{(C_1\xi_{2})^2}\sim 10^{-10}$. Another
feature of this model is that the SM Higgs field is free from the
non-minimal coupling of gravity.

\section{Conclusions}

In this work, we have addressed the inflation in the framework of the SM
with the presence of 4-form interactions.
Compared to the existing models of Higgs inflation, we have modified
the values of $\lambda$ and $\xi$ in the presence of
 4-form-flux couplings.
 We also see that the four-form interactions can
  generate the Higgs-boson mass before electroweak symmetry breaking
  and such generated Higgs mass is also quantized.
It is suggested that the 4-form flux plays an important role in this
inflationary
scenario. By switching off the components of
$\xi_{eff}$, we can re-obtain the minimal and non-minimal Higgs inflation
scenarios.

We have introduced
the most general Lagrangian for the inflation models.
We have also realized that there is no need for the Higgs-gravity
coupling to produce the right amount of density perturbation.
In Ref.~\cite{a} the authors discussed about the requirement of
the Higgs-gravity coupling in order to give 
a successful inflation model. However, we have shown explicitly that
it is possible to address the inflation without the Higgs
non-minimal coupling.  We can reproduce the results of
Ref.~\cite{a} from our model.

\section*{Acknowledgment}
This work was supported by the Taiwan Ministry of Science and Technology
under Grant No. MOST-107-2112-M-007-029-MY3.

\end{document}